\documentclass[a4paper,twocolumn,11pt,accepted=2022-06-07]{quantumarticle}
\pdfoutput=1
\usepackage[utf8]{inputenc}
\usepackage[english]{babel}
\usepackage[T1]{fontenc}
\usepackage[numbers]{natbib}
\usepackage{listings}

\usepackage{amsmath}
\usepackage{hyperref}
\usepackage{tikz}
\usepackage[resetlabels]{multibib}

\usepackage{mathrsfs}
\usepackage{graphicx}
\usepackage{epstopdf}
\usepackage{dcolumn}
\usepackage{bm}
\usepackage{color} 
\usepackage{CJK}
\usepackage{lipsum}

\newcites{article}{article references}
\newcites{book}{book references}
\newcites{misc}{misc references}
\newcites{repo}{repository references}
\newcites{web}{website references}
\newcites{other}{Other references}

\def\la{{\langle}}
\def\ra{{\rangle}}

\newcommand{\beq}{\begin{equation}}
\newcommand{\eeq}{\end{equation}}
\newcommand{\beqa}{\begin{eqnarray}}
\newcommand{\eeqa}{\end{eqnarray}}

\begin{document}
\title{Inverse engineering of fast state transfer among coupled oscillators}

\author{Xiao-Jing Lu}
\affiliation{School of Science, Xuchang University, Xuchang 461000, China}
\affiliation{Departamento de Qu\'{\i}mica F\'{\i}sica, University of the Basque Country UPV/EHU, Apdo.
644, 48080 Bilbao, Spain}
\email{luxiaojing1013@163.com}
\orcid{0000-0002-7887-0159}

\author{Ion Lizuain}
\affiliation{Department of Applied Mathematics, University of the Basque Country UPV/EHU, Donostia-San Sebasti\'an, Spain}
\affiliation{EHU Quantum Center, University of the Basque Country UPV/EHU}
\orcid{0000-0001-9207-4493}

\author{J. G. Muga}
\affiliation{Departamento de Qu\'{\i}mica F\'{\i}sica, University of the Basque Country UPV/EHU, Apdo.
644, 48080 Bilbao, Spain}
\affiliation{EHU Quantum Center, University of the Basque Country UPV/EHU}
\orcid{0000-0002-1967-502X}

\maketitle

\begin{abstract}
We design faster-than-adiabatic state transfers (switching of quantum numbers)
in  time-dependent coupled-oscillator Hamiltonians.
The manipulation to drive the process is found using
a two-dimensional invariant recently proposed in  {\it S. Simsek and F. Mintert, Quantum 5 (2021) 409}, and involves
both rotation and transient scaling of the principal axes of the potential in a Cartesian representation.
Importantly, this invariant is degenerate except for the
subspace spanned by its ground state. Such degeneracy, in general, allows for infidelities of the final states with respect to ideal target eigenstates. However, the value of a single control parameter
can be chosen so that  the state switching is perfect  for arbitrary (not necessarily known) initial eigenstates. Additional
2D linear invariants are used to find easily the parameter values needed and to
provide generic expressions for the final states and final energies.
In particular we find
time-dependent transformations of  a two-dimensional harmonic trap for a particle (such as an ion or neutral atom)
so that the final trap is  rotated with respect to the initial one, and eigenstates of the initial trap are converted into rotated replicas
at final time,  in some chosen time and rotation angle.

\end{abstract}
%

\section{introduction}
\begin{figure}[tbp]
~~~~~~~~~~~\scalebox{0.54}[0.54]{\includegraphics{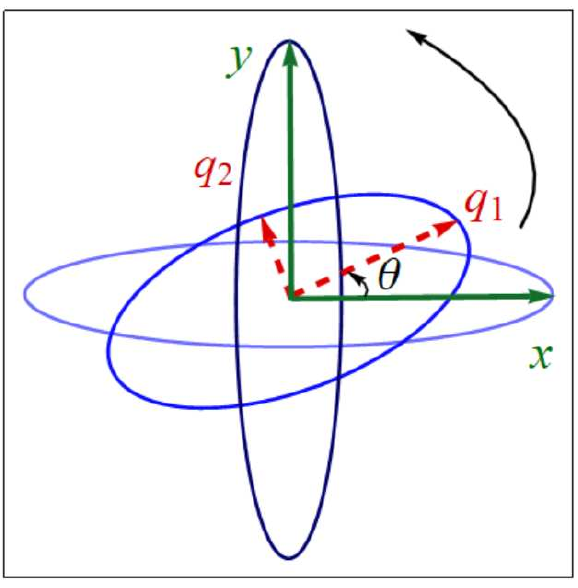}}
\caption{(Color online) Schematic representation of initial (long principal axis in $x$ direction) and final (long principal axis in $y$ direction)
configurations of a 2D harmonic potential
as well as an intermediate configuration with principal axes along $q_1$ and $q_2$.}\label{fig0}
\end{figure}

A goal of current Physics for fundamental science and technological applications such as metrology, sensing,  or information processing is to
achieve a better control of states and dynamics of single or few-body quantum systems
such as ions or neutral atoms.  Both internal and motional states have to be controlled. We shall focus here on setting fast control
protocols for two-dimensional (2D) systems
described by Hamiltonians of the (dimensionless) form
\beq
H(t)=\frac{p^2_x}{2}+\frac{p^2_y}{2}+
\frac{1}{2}M_{11}x^2+\frac{1}{2}M_{22}y^2
+M_{12}xy,\label{H1}
\eeq
with $M$ real, symmetric ($M_{12}=M_{21}$), and controllable in time.
More specifically,
we wish to consider a faster than adiabatic process  to switch eigenstates among the $x$ and $y$ oscillators, namely,
$|n,k\ra_{i} \to |k,n\ra_{f}$, up to a phase factor,  (where $i$ and $f$ stand for initial and final configurations) for arbitrary
vibrational quantum numbers $n$, $k$ and the uncoupling condition $M_{12}=0$ at boundary times.
This Hamiltonian may represent in particular two oscillators on a line transiently coupled, or one particle in a time-dependent 2D trap, see other possible interpretations in ref. \cite{Tobalina2020}.
Since the different systems are mathematically equivalent and the single particle system is easier to visualize we shall from now on
use a language appropriate for the manipulation of a Cartesian 2D potential that holds a single particle.
Specifically
we shall look for a process  where the final potential is rotated
by $\pi/2$ with respect to the initial one, see Fig. \ref{fig0}.
This implies the additional boundary relations  $(M_{22})_f=(M_{11})_i$ and
$(M_{11})_f=(M_{22})_i$. Other natural boundary conditions such as $(M_{11})_f=(M_{11})_i$ and
$(M_{22})_f=(M_{22})_i$, or rotations by a different rotation angle will be also discussed later.

Faster-than-adiabatic, excitationless  rotations \cite{Masuda2015,Masuda2011,Palmero2016,Lizuain2019,Lizuain2017}, in particular, form, together
with transport  \cite{Palmero2013,Palmero2014,Lu2018,Lu2014}, expansions/compressions \cite{Chen2010a,Palmero2015} or separations/mergings \cite{Martinez-Garaot2018,Rodriguez-Prieto2020},  a basic set of motional operations. These operations are needed to make the architectures based on moving the qubits encoded in ions or neutral atoms viable for practical implementations of information processing \cite{Kielpinski2002}.
Fast rotations have been performed with increasing accuracy experimentally but remain technically challenging \cite{Splatt2009,Kaufman2017,Urban2019,vanMourik2020}. From the theory side, they also imply some difficulties, even for the simple setting of a rotating harmonic trap. In essence, a point  transformation (i.e., one that does not mix positions and momenta)
cannot separate the dynamics into independent normal dynamical modes, so that simple inverse-engineering methods known for the one-dimensional modes and the corresponding 1D invariants cannot be used to find a fast, excitationless rotation protocol \cite{Lizuain2017}.
Several wayouts have been investigated, such as purely numerical optimizations \cite{Tobalina2021}, or mode separations using  more complicated transformations mixing positions and momenta \cite{Lizuain2019}. However, numerical optimizations need specific calculations for each particular
state and process,
and the approach based on non-point transformations was limited by singularities in the protocol and conditions on the rotation times and commensurate normal frequencies \cite{Lizuain2019}. Also, for a charged particle,  if the  trap manipulation is complemented with a time-dependent magnetic field
perpendicular to the rotation plane, and a  time-dependent scaling of the principal axes to compensate for non-inertial effects, fast rotations can be done
without final excitation \cite{Masuda2015,Masuda2011,Lizuain2017}. We assume here that such compensation is not viable, either because the Hamiltonian does not represent a charged particle  or because of technical limitations. While these mentioned wayouts might be effective solutions in some cases, the need for a robust methodology to design smooth, fast trap and state rotations,
remains. Here ``robustness" refers, in the spirit of simple  shortcut-to-adiabaticity (STA) processes \cite{Guery2019}, to independence with respect to the initial state.
In this work we look for, and find, such a robust methodology. We shall make use of recent work on two-dimensional invariants \cite{Tobalina2020,Simsek2021}. First we adapt to the rotation scenario a recently proposed two-dimensional quadratic invariant commuting with initial and final trap Hamiltonians
\cite{Simsek2021} to define the
time-dependent protocols for the harmonic trap. Inverse engineering will proceed from the 2D invariant to the evolution of the 2D harmonic potential, which
will include both rotations and scaling of the instantaneous eigenfrequencies.  Because of the  degeneracy of this invariant, an arbitrary initial eigenstate of the Hamiltonian leads to a combination of states in the degenerate subspace which in general does not conserve the initial energy at final time.\footnote{
The solutions of the Schr\"odinger equation for a time dependent Hamiltonian may be written in terms of orthonormal eigenstates
$|\Phi_{{N,\alpha}};t\ra$  of a Hermitian invariant,
where $N$ is the eigenvalue and $\alpha$ a label to distinguish orthogonal eigenstates in the degenerate subspace, as \cite{Lewis1969}
$
|\psi(t)\ra=\sum_N \sum_\alpha  a_{_{N\alpha}} e^{i\Theta_{_{\!N\alpha}}\!(t)} |\Phi_{N,\alpha}; t\ra,
$
where the $a_{_{N\alpha}}$ do not depend on time and $\Theta_{_{\!N\alpha}}\!(t)$ is the Lewis-Riesenfeld phase, defined so that
$e^{i\Theta_{_{\!N\alpha}}\!(t)} |\Phi_{N,\alpha}; t\ra$ is a solution of the Schr\"odinger equation.
} (An exception is the non-degenerate subspace of the ground state.)

To solve this problem we construct auxiliary linear invariants following Tobalina et al. \cite{Tobalina2020}. They will enable us to find the final state, as well
as the energy in terms of only a pair of complex-valued classical trajectories. We can thus easily find parameters for which the process performs as a perfect rotation for any (not necessarily known) initial eigenstate.

All numerical calculations of the dynamics are performed  in several ways in this paper:
The Wigner function provides a useful tool since it evolves classically in a harmonic trap; we independently test
the results by means of a split-operator, fast-Fourier transform approach, and also using a finite basis in a moving frame.
Appendix \ref{sectionwig} gives details on the Wigner function method.

In Section \ref{sectionframe}
we introduce the notation and Hamiltonian; in Section \ref{sectioninvariant}
we describe the  two-dimensional invariant adapted from Simsek et al. \cite{Simsek2021}; in Section \ref{sectionmodel}, the potential is designed based on the invariant, and in Section \ref{deg} the degeneracy of the invariant is discussed;
In Section \ref{otherinv}, auxiliary linear invariants are introduced; Section \ref{fs} uses the different invariants to provide a generic expression
of the final state, and it describes how to achieve perfect state transfers solving the degeneracy problem.
The article ends with a Discussion and technical Appendices.
\section{Notation and model}\label{sectionframe}
As a relevant example of a physical setting that leads to the Hamiltonian  (\ref{H1}) we consider a two dimensional space of (laboratory) fixed axes $\tilde{x},\tilde{y}$ and a particle of mass $m$ moving
in an anisotropic harmonic potential
whose principal  axes $\tilde{q}_1, \tilde{q}_2$ rotate by $\theta$ (counterclockwise polar angle measured from the $\tilde{x}$ axis to the $\tilde{q}_1$ axis)
and admits as well a time dependence of the instantaneous eigenfrequencies $\tilde{\omega}_{1,2}(t)$.
From now on we shall  consider dimensionless versions  (without tilde) of the corresponding coordinates, defined  in terms of a unit of position $\sqrt{\hbar/(m\tilde{\omega}_1(0))}$,
unit of momentum $\sqrt{\hbar m\tilde{\omega}_1(0)}$, unit of time $1/\tilde{\omega}_1(0)$, and unit of energy $\hbar \tilde{\omega}_1(0)$.
The coordinate and momentum transformations between laboratory and rotating coordinates and momenta  are
%
%
%
%

\beq
\left(\begin{array}{cc}
              {q}_1  \\
               {q}_2 \\
             \end{array}
           \right)=W(t)\left(\begin{array}{cc}
              x  \\
              y \\
             \end{array}
           \right),~~
           \left(\begin{array}{cc}
              {p}_1  \\
              {p}_2 \\
             \end{array}
           \right)=W(t)\left(\begin{array}{cc}
             p_x  \\
              p_y \\
             \end{array}
           \right)
\label{eq2}
\eeq
with
\beqa
W(t)=\left(
       \begin{array}{cc}
         \cos\theta & \sin\theta \\
         -\sin\theta & \cos\theta \\
       \end{array}
     \right).
\eeqa
Adding kinetic and potential terms the lab-frame Hamiltonian may simply be written as
\beqa
H(t)&=&\frac{p^2_x}{2}+\frac{p^2_y}{2}+\frac{1}{2}{\omega}^2_1(t)q_1^2(x,y)
\nonumber\\
&+&\frac{1}{2}{\omega}^2_2(t)q_2^2(x,y)
\nonumber\\
&=&\frac{p^2_x}{2}+\frac{p^2_y}{2}+\frac{1}{2}{\omega}^2_1(t)(x \cos\theta+y\sin\theta)^2
\nonumber\\
&+&\frac{1}{2}{\omega}^2_2(t)(x \sin\theta-y\cos\theta)^2,
\label{Hlab}
\eeqa
%
which is of the form (\ref{H1}) with
%
%
%
\beqa
M_{11}&=&{\omega}^2_1\cos^2\theta+{\omega}^2_2\sin^2\theta,\\
M_{22}&=&{\omega}^2_1\sin^2\theta+{\omega}^2_2\cos^2\theta,\\
M_{12}&=&M_{21}=({\omega}^2_1-{\omega}^2_2)\sin\theta\cos\theta,
\eeqa
and
\beqa
\tan(2\theta)&=&\frac{M_{12}+M_{21}}{M_{11}-M_{22}},
\nonumber\\
{\omega}^2_1&=&\frac{\cos^2\theta}{\cos2\theta}M_{11}-\frac{\sin^2\theta}{\cos2\theta}M_{22},
\nonumber\\
{\omega}^2_2&=&-\frac{\sin^2\theta}{\cos2\theta}M_{11}+\frac{\cos^2\theta}{\cos2\theta}M_{22}.
\eeqa
\section{The primary invariant to design the harmonic trap evolution}\label{sectioninvariant}
%
%
%
%
\begin{figure}[tbp]
\scalebox{0.42}[0.42]{\includegraphics{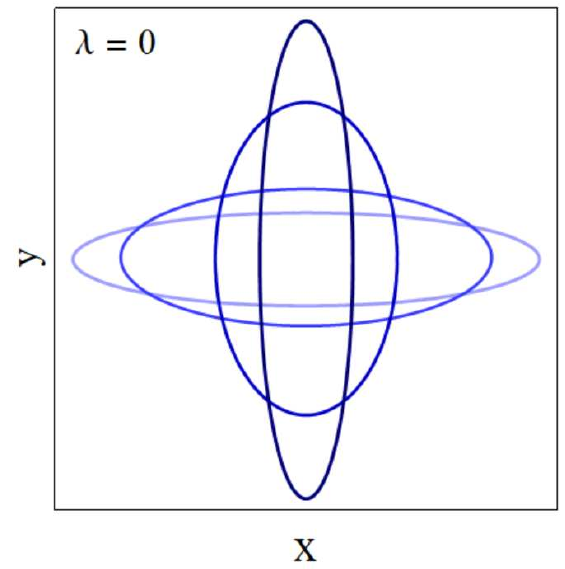}}
\scalebox{0.42}[0.42]{\includegraphics{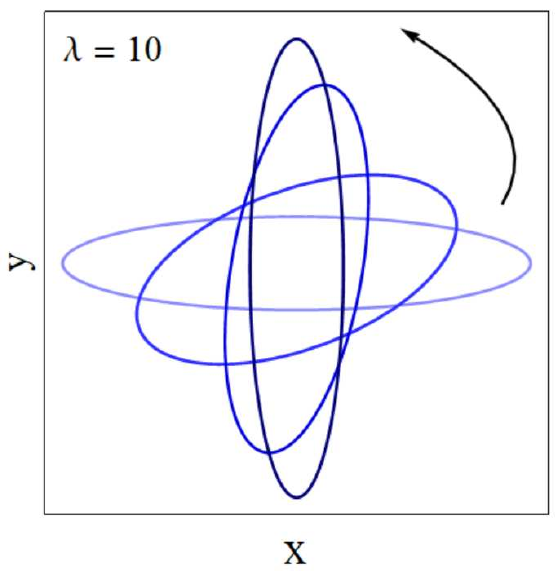}}
\scalebox{0.41}[0.41]{\includegraphics{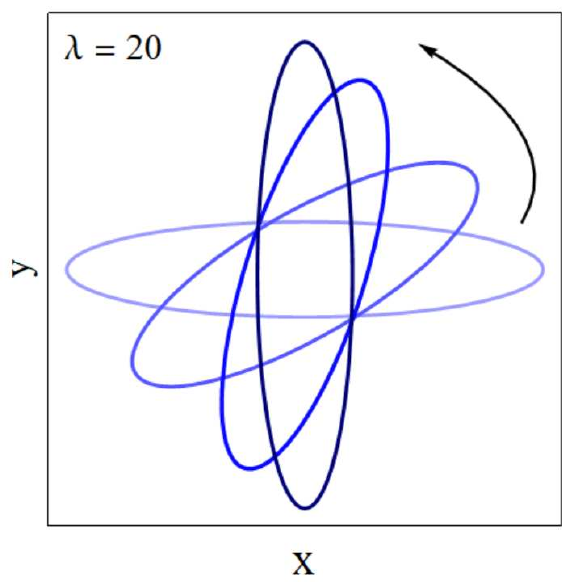}}
\scalebox{0.41}[0.41]{\includegraphics{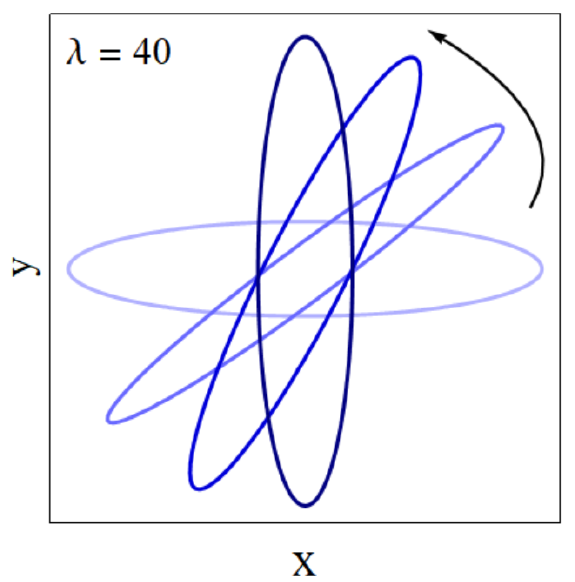}}
\caption{(Color online) Snapshots at equal time intervals of equipotential lines (time increases from lighter to darker lines) of the trap for different values of the control parameter $\lambda$. For $\lambda=0$ the trap simply expands in $y$-direction and compresses in $x$-direction. For $\lambda\ne 0$ the trap rotates anticlockwise with some deformation of the principal axes. In all cases $x$ and $y$  initial and final frequencies are switched. Parameters: $t_f=5$, $w_2=5$.}\label{fig1}
\end{figure}
%
%
\begin{figure}[tbp]
\scalebox{0.8}[0.8]{\includegraphics{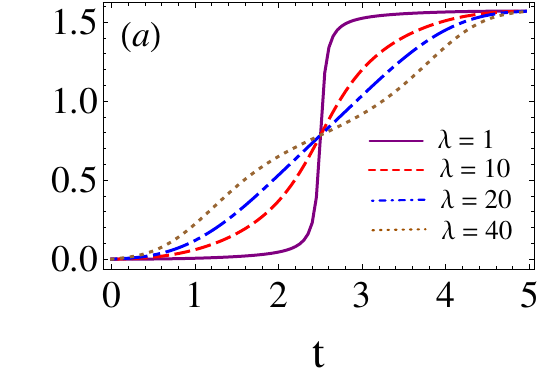}}
\scalebox{0.8}[0.8]{\includegraphics{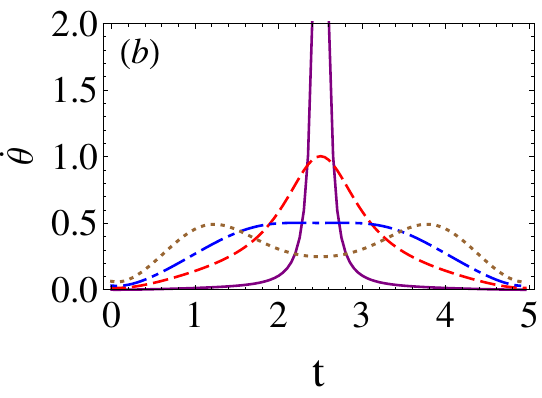}}
\caption{(Color online) $\theta$ and $\dot{\theta}$ versus time for  $\lambda=1$ (purple solid line), $\lambda=10$ (red short dashed line), $\lambda=20$ (blue dot-dashed line) and $\lambda=40$ (brown dotted line). The peak cut in (b) reaches $10$. $t_f=5$, $w_2=5$.}\label{figtheta}
\end{figure}
\begin{figure}[tbp]
	\scalebox{0.8}[0.8]{\includegraphics{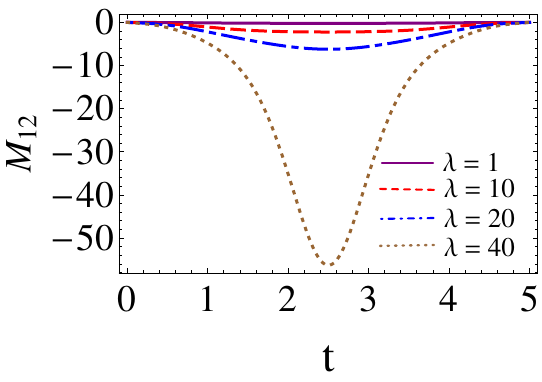}}
	\caption{(Color online) Coupling term $M_{12}(t)$ for different $\lambda$.  $t_f=5$, $w_2=5$.}\label{figm12}
\end{figure}
The Hamiltonian (\ref{H1}) can be written as a $4\times4$   matrix
\beqa
H=\frac{1}{2}\hat{X}^T\Omega\hat{X},~~~
\Omega=\left(
             \begin{array}{cc}
                M &0 \\
               0 & \mathbf{1} \\
             \end{array}
           \right),
           \label{H2}
\eeqa
where $\hat{X}^T=(x,y,p_x,p_y)$,  $\mathbf{1}$ represents a $2\times2$ unit matrix, and $M$ is a time-dependent (potential) matrix
\beqa
M=\left(
             \begin{array}{cc}
              M_{11} &M_{12} \\
               M_{21} &M_{22} \\
             \end{array}
           \right).
\eeqa
So far the formal results are valid for classical or quantum particles. For a quantum particle the components of $\hat{X}$ are  to be interpreted as operators with the standard position/momentum commutation relations. In the following, we will construct the invariant to design the rotation following ref. \cite{Simsek2021},  in a quantum frame.
To find an invariant,
a quadratic ansatz may be assumed with the form
\beq
I_0(t)=\frac{1}{2}\hat{X}^T{\Gamma}\hat{X},
\label{I0}
\eeq
where the subscript $0$ will distinguish it from other invariants to be defined later.
The invariant $I_0$ has to satisfy the relation of all invariants $\partial I(t)/\partial t=i [I(t),H(t)]$, which implies
\beq
\frac{d{\Gamma}}{dt}=\Omega \mathcal{S} \Gamma-\Gamma \mathcal{S} \Omega,~~~
\mathcal{S}=\left(
             \begin{array}{cc}
               0 &{\bf 1} \\
               -{\bf 1} & 0 \\
             \end{array}
           \right).\label{f1}
\eeq
As invariants imply specific dynamics, $H(t)$ may be inverse engineered from $I_0(t)$, but care must be exercised,
as the generic form (\ref{I0}) might lead to a $H(t)$  without the desired form (\ref{H2}).  To impose the
desired form,  $\Gamma$ is designed with the parameterization \cite{Simsek2021}
\beqa
\Gamma\!=\Re\left(
             \begin{array}{cc}
               \dot{P}^{\dagger}\dot{P} &-\dot{P}^{\dagger}P \\
               -{P}^{\dagger}\dot{P} & P^{\dagger}{P} \\
             \end{array}
           \right),\label{f2}
\eeqa
where $P$ is a $2\times2$ complex matrix, $\Re$ stands for the real part of the matrix, and the dots are time derivatives.
Using Eq. (\ref{f2}) into Eq. (\ref{f1}),
\beqa
\left(\begin{array}{cc}
               \Re(\dot{P}^{\dagger}{\cal D}+{\cal D}^{\dag}\dot{P}) &-\Re({\cal D}^{\dagger}P )\\
               -\Re(P^{\dagger}{\cal D}) & 0 \\
             \end{array}
           \right)=0,
\label{conditionHI}
\eeqa
where ${\cal D}=\ddot{P}+P M$. Any $P$ such that  ${\cal D}=0$ leads to an invariant consistent with  Eq. (\ref{H2}), but additional restrictions are necessary to make $M$ real-symmetric.
To that end it is useful to assume a polar decomposition $P=U R$ with $U$ unitary and $R$ Hermitian and positive-semidefinite.   $R$ is left as a free matrix and $U$ is determined to make $M$ real-symmetric by defining
$A=U^{\dagger}\dot{U}$, which gives $U$ from $A$ and the initial condition $U(0)=1$.
Hermiticity of $M$ requires
\beqa
A=i R^{-2}+\frac{1}{2}[R^{-1},\dot{R}]+\frac{1}{2}R^{-1}{\cal J} R^{-1},
\nonumber\\
\{{\cal J},R^{-2}\}=[\dot{R},R^{-1}]+[R,R^{-2}]_{\dot{R}},
\label{aeq}
\eeqa
where  $[x,y]_z=xzy-yzx$  and $\{x,y\}=xy+yx$ is the anti-commutator.
%
One then finds the differential equation for $R$ \cite{Simsek2021},
\beq
\{\ddot{R},R\}+\{R^2,M\}=2[\dot{R},R]_A-2RA^2R,
\label{M}
\eeq
which guarantees that $M$ is real-symmetric for all times if $R$ is real and positive and  $\dot{R}(0)=0$ \cite{Simsek2021}. Based on Eqs. (\ref{aeq}) and (\ref{M}),
the Hamiltonian is inverse engineered by designing  $R$ with proper boundary conditions, and then
using Eqs. (\ref{aeq}) and (\ref{M}). An explicit expression for $M$ in terms of $R$ may be found in Eq. (79) of ref. \cite{Simsek2021}.
The boundary conditions are chosen so that
\beq
[I_0(0),H(0)]=[I_0(t_f),H(t_f)]=0,
\label{commut}
\eeq
which implies
\beqa
\ddot{R}(0)=\dot{R}(0)=0,~~~\ddot{R}(t_f)=\dot{R}(t_f)=0,\label{Rdc}\\
R(0)=M(0)^{-\frac{1}{4}},~~~~R(t_f)=M(t_f)^{-\frac{1}{4}}.\label{Rc}
\eeqa
In simple scenarios without degeneracies of $H$ and $I$, the commutativity at the boundaries (\ref{commut}) implies a one-to-one dynamical mapping from
initial to final eigenstates of the Hamiltonian, see e.g. \cite{Chen2010a}, carried out by the eigenstates of the invariant. This is the basis for
many inverse engineering applications from $I(t)$ (representing the desired dynamics)  to $H(t)$ (the  driving) \cite{Guery2019}.
The current $I_0$, however, is a degenerate operator, see below, which makes the desired mapping possible,  as we shall see, but not necessary.
A basic goal of this work is to find out the way to impose that mapping.
%
%
%
\begin{figure}[tbp]
\scalebox{0.7}[0.7]{\includegraphics{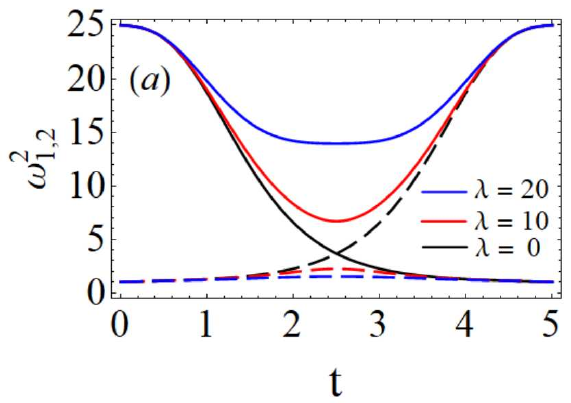}}
\scalebox{0.7}[0.7]{\includegraphics{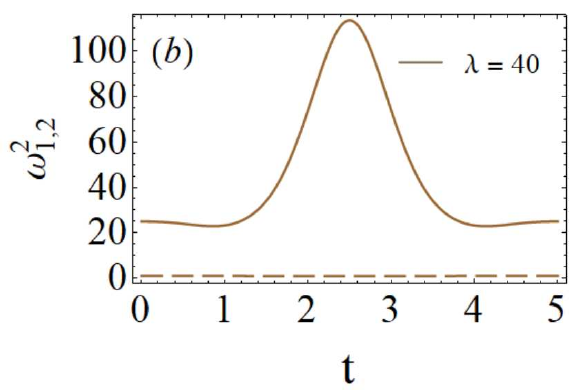}}
	\caption{(Color online) Trap frequencies squared ${\omega}^2_{1,2}$  versus time for different $\lambda$, the dashed line is for ${\omega}_1^2$ and
		the solid line for ${\omega}_2$.
		The lines do cross at $\lambda=0$ but the crossing is avoided for $\lambda\neq 0$. $t_f=5$, $w_2=5$.
	}\label{figomega}
\end{figure}
\section{Design of $H(t)$ from the invariant}\label{sectionmodel}
We consider the following
initial and final potentials
\beqa
V_i=\frac{1}{2} w^2_1 x^2+\frac{1}{2} w^2_2 y^2,
\nonumber\\
V_f=\frac{1}{2} w^2_2 x^2+\frac{1}{2} w^2_1 y^2,
\label{vbound}
\eeqa
where the initial and final frequencies in $x$ and $y$ directions swap,
and we use a special notation for initial and final frequencies, namely,
$w_1\equiv\omega_1(0)=\omega_2(t_f)$ and  $w_2=\omega_2(0)=\omega_1(t_f)$.
Let us recall that with our units $w_1=1$ but in several equations we
write $w_1$ (instead of its value 1) to achieve a more clear physical interpretation.
Our goal is to transform the potential from $V_i$ to $V_f$ so that initial eigenstates
$|n,k\ra_i$  become their rotated versions at final time, namely $|k,n\ra_f$, up to a phase factor.
An adiabatic rotation of the principal axes would accomplish such transformations but we want faster
drivings using the invariant as a guidance for inverse engineering $H(t)$.
From the boundary conditions in Eqs. (\ref{vbound}) and  (\ref{Rc}), $R$ has to satisfy
\beqa
R(0)\!\!=\!\!\left(
  \begin{array}{cc}
    \!\!w^{-\frac{1}{2}}_1\!\! & \!\!0\!\! \\
    \!\!0\!\! & \!\!w^{-\frac{1}{2}}_2\!\! \\
  \end{array}
\right),~
R(t_f)\!\!=\!\!\left(
  \begin{array}{cc}
    \!\!w^{-\frac{1}{2}}_2\!\! & \!\!0\!\! \\
    \!\!0\!\! & \!\!w^{-\frac{1}{2}}_1\!\! \\
  \end{array}
\right).\nonumber\\
\label{Rt}
\eeqa
(The exponents of $R$ are instead $+1/2$  in  \cite{Simsek2021}, which is a typo.)

Considering the conditions in Eqs. (\ref{Rdc}) and (\ref{Rc}), $R(t)$  is interpolated with a polynomial $p(t)$,
\beqa
R(t)&=&[1-p(t)]R(0)+\!p(t)R(t_f)\nonumber\\
&+&(t/t_f)^3(1-t/t_f)^3R_c,
\eeqa
where $R_c$ is in principle an arbitrary real symmetric matrix and\footnote{
These forms of $p(t)$ and $R(t)$ are just a simple choice and none of the formal results to be found later on depend on them. 
Alternative forms may be used as long as the boundary conditions are satisfied and R(t) remains positive semidefinite, 
for example, to optimize robustness with respect to perturbations, or other variables such as transient energies\cite{Espinos2022}.}
\beq
p(t)=10(t/t_f)^3-15(t/t_f)^4+6(t/t_f)^5,
\eeq
so that the boundary conditions (\ref{Rdc}) are also satisfied.
%
%
A purely diagonal $R_c$ implies that the principal axes do not rotate.
This would lead to a mapping of the form $|n,k\ra_i\to |n,k\ra_f$ instead of the desired $|n,k\ra_i\to |k,n\ra_f$ (up to phase factors).
Thus we choose instead
\beqa
R_c=\lambda(w_1 w_2)^{-1/4}\left(
  \begin{array}{cc}
    0 & 1 \\
    1 & 0 \\
  \end{array}
\right)
\eeqa
with $\lambda$ as a free parameter. Figure \ref{fig1} shows snapshots of the trap (in the laboratory frame) at equal time intervals between $t=0$ and $t_f$ for different  $\lambda$.
For $\lambda=0$ there is no rotation but a pure compression (in $x$ direction) and expansion (in $y$ direction)
of the two orthogonal oscillators with principal axes fixed in the laboratory frame. For $\lambda\ne 0$
the principal axes rotate, more abruptly as $\lambda\to 0$, see Fig. \ref{figtheta}. For increasing values of $\lambda$ the rotation velocity shows
first a maximum at $t_f/2$ but eventually develops two maxima.
The effect of $\lambda$ on the coupling term $M_{12}$  is depicted in Fig. \ref{figm12}, and on
$\omega^2_{1,2}(t)$ in Fig. \ref{figomega}.

\begin{table*}[t]
\centering
\begin{tabular}{||c c c c c c c c c c c ||}
\hline
\!\!Initial state \!\!&\!\!$|0,0\ra_i$\!\!&$|1,1\ra_i$&\!\!$|1,0\ra_i$&\!\!$|0,1\ra_i$\!\!&\!\!$|2,0\ra_i$&\!\!$|0,2\ra_i$\!\!&$|3,0\ra_i$\!\!&$|0,3\ra_i$\!\!&$|2,1\ra_i$\!\!&$|1,2\ra_i$
\\\hline
$E_{nk}(t=0)$         \!\! &\!\!3    \!\!  &9      \!\!     &4   \!\!   &8         &5      &13       &6      &18      &10     &14        
\\\hline
$\delta$       \!\! &\!\!0    \! \! &0   \!\!   &2.613  \!\! &-2.613   &5.227   &-5.227  &7.840   &-7.840  &2.613  &-2.613 
\\\hline
\end{tabular}
\caption{For a given initial state $|n,k\ra_i$,
Initial energy $E_{nk}(0)=(n+1/2)w_1+(k+1/2)w_2$ and final excess energy $\delta=\langle H(t_f)\rangle-E_{nk}(0)$. Parameters: $t_f=1$, $\lambda=20$, $w_2=5$.}\label{table2}
\end{table*}
%
%
%
%
\section{Degeneracy of $I_0$\label{deg}}
The invariant may be calculated at initial and final times
from the boundary conditions. In terms of number operators for $x$ and $y$ oscillators
it takes the form
\beq
I_0(t_b)=n_x(t_b)+n_y(t_b)+1,
\eeq
where $t_b=0,t_f$.
For simplicity it is useful to subtract the constant and define $I_+=I_0-1$ as a pure sum of number operators at boundary times,
\beq
I_+(t_b)=n_x(t_b)+n_y(t_b).
\eeq
Being an invariant, the spectrum of $I_+(t)$ is  constant, and  degenerate at all times, except for the ground state.
The different subspaces can be labelled by $N=0,1,2...$, where $N$ is the sum of quantum numbers, and their dimension is
$N+1$. The non-degenerate subspace $N=0$ is spanned by the ground states of the Hamiltonian at boundary times.
At any other time the state that dynamically evolves from $|0,0\ra_i$ is the nondegenerate ground eigenstate of $I_+(t)$.
In general it is not an eigenstate of $H(t)$ except in the limit of very slow, adiabatic processes.
The next subspace is  $N=1$, spanned by $|0,1\ra_i$ and $|1,0\ra_i$ at initial time and by $|0,1\ra_f$ and $|1,0\ra_f$ at final time. The states that evolve dynamically from $|0,1\ra_i$ and $|1,0\ra_i$ remain orthogonal, are degenerate eigenstates of $I_+(t)$, and span at any time the
subspace $N=1$ of the invariant. However there is in general no guarantee that the state $|0,1\ra_i$ will become
$|1,0\ra_f$. A priori, all we can say is that it will evolve into a linear combination of $|0,1\ra_f$ and $|1,0\ra_f$ at $t_f$.
As a further example,
$N=2$ is spanned at initial and final time by $|0,2\ra_{i,f}$, $|2,0\ra_{i,f}$,  and $|1,1\ra_{i,f}$; and so on. Any state $|n,k\ra_i$ evolves in time within the
invariant subspace $n+k$, which is spanned by the states that evolve dynamically from the initial eigenstates but, except for the ground state,
the final state is in general a linear combination of the eigenstates of $H(t_f)$ that span the subspace, rather than the desired
$|k,n\ra_f$. Numerical calculations, see an example in Table I,  demonstrate that the energy increment when the state starts at $|n,k\ra_i$ takes a   remarkably simple form, namely,\footnote{We found this expression of the final energy first heuristically from the calculations using the Wigner function for different values of $n$ and $k$ as exemplified in Table I. Later on we could find an analytical expression for the time dependent wavefunction, see Eq. (\ref{finalstate}) below,
that allows to calculate the final energy analytically for given a given pair $k,n$, as shown in the Appendix.
This also gives an expression for $b$.}
\beq
\la H(t_f)\ra-E_{nk}(0)=b (n-k)(w_2-w_1),
\label{increment}
\eeq
where $E_{nk}(0)=E_{kn}(t_f)=(n+1/2)w_1+(k+1/2)w_2$, and the proportionality
constant $b$ depends, for given $w_{1,2}$,  only on the process time $t_f$, and on $\lambda$, but not on $n$ or $k$.
This might at first sight seem to indicate that at least $|n,n\ra_i,\,\, n>0$ becomes
$|n,n\ra_f$, but that is not the case in general as we shall see.
In fact we shall discuss later the physical meaning  of the constant $b$, which, for a given $t_f$, can be manipulated and made
zero by choosing the proper value of $\lambda$. We shall also see that this choice corresponds indeed to the desired
process $|n,k\ra_i\to|k,n\ra_f$, for all $n,k$.
To proceed further we need to  introduce  more invariants.

\section{Construction of further invariants\label{otherinv}}

To get a more explicit expression for the final state and final energy for the processes designed from $I_+(t)$,
we shall make use of further invariants.
The linear operators \cite{Castanos1994,Urzua2019,Tobalina2020}
\beq
G(t)=u_x(t)p_x-\dot{u}_x(t)x+u_y(t)p_y-\dot{u}_y(t)y
\eeq
are invariants provided
\beqa
\ddot{u}_x+M_{11}(t)u_x&=&-M_{12}(t)u_y,\nonumber\\
\ddot{u}_y+M_{22}(t)u_y&=&-M_{21}(t)u_x,\label{u-equation}
\eeqa
which are classical (Newton) equations of motion driven by a Hamiltonian (\ref{H1}).
We may thus regard  $u_x(t),u_y(t)$ as the Cartesian components of a classical trajectory.
Note that the $u_x,u_y$ might be complex, representing in that case two trajectories,
since real and imaginary parts evolve independently and solve the classical equations.
It proves useful to write $G(t_b)$ at boundary times $t_b=0,t_f$ in terms of
(initial and final) creation and annihilation operators,
\beqa
G(t_b)
&=&\!\sum_{z=x,y}\frac{a^{\dag}_z(t_b)}{\sqrt{2}}\bigg[i\sqrt{\omega_z(t_b)}u_z(t_b)-\frac{\dot{u}_z(t_b)}{\sqrt{\omega_z(t_b)}}\bigg]
\nonumber\\
&-&\!\!\sum_{z=x,y}\frac{a_z(t_b)}{\sqrt{2}}\bigg[\!i\sqrt{\omega_z(t_b)}u_z(t_b)\!+\!\frac{\dot{u}_z(t_b)}{\sqrt{\omega_z(t_b)}}\!\bigg],\nonumber\\
\label{G}
\eeqa
where $a_z(t_b)=\sqrt{\omega_z(t_b)/2}z+i p_z/\sqrt{2\omega_z(t_b)}$, $z=x,y$.
Different linear invariants may be  constructed by choosing specific boundary conditions for $u_{x,y}$ and $\dot{u}_{x,y}$.
In particular,
the initial conditions \cite{Tobalina2020}
\beqa
u_x(0)&=&i/\sqrt{2\omega_x(0)},~~\dot{u}_x(0)=-\sqrt{{\omega_x(0)}/{2}},
\nonumber\\
u_y(0)&=&0,~~~~~~~~~~~~~~~\dot{u}_y(0)=0,
\eeqa
define an invariant which is initially $G_1(0)=a_x(0)$.
Instead, the initial conditions
\beqa
&&\!\!u'_x(0)\!=\!0,~~~~~~~~~~~~~\dot{u}'_x(0)\!\!=\!0,\nonumber\\
&&\!\!u'_y(0)\!\!=\!i/\sqrt{2\omega_y(0)},~~\dot{u}'_y(0)\!\!=\!\!-\sqrt{{\omega_y(0)}/{2}},
\label{initial-ux}
\eeqa
define a different invariant which at time zero is $G_2(0)=a_y(0)$. Here we use the prime $'$ to distinguish the trajectories with initial
conditions in Eq. (\ref{initial-ux}).

We may construct corresponding quadratic invariants as $I_1(0)=G_1^\dagger(0) G_1(0)$ and $I_2(0)=G_2^\dagger(0) G_2(0)$.
Clearly $I_+(0)=n_x(0)+n_y(0)=I_1(0)+I_2(0)$ so, for the $H(t)$ inverse engineered from $I_+(t)$,  we have that $G_1^\dagger(t_f)G_1(t_f)+G_2^\dagger(t_f) G_2(t_f)=I_+(t_f)=
n_x(t_f)+n_y(t_f)$. This implies that $G_1(t_f)$ and $G_2(t_f)$ have no $a_z^\dagger(t_f)$ components and
the following  relations hold for the final values of the trajectories, see Eq. (\ref{G}),
\beqa
&&\dot{u}_x(t_f)\!=\!i\omega_x(t_f)u_x(t_f),\,\,
\dot{u}_y(t_f)\!=\!i\omega_y(t_f)u_y(t_f),
\nonumber\\
&&\dot{u}'_x(t_f)\!=\!i\omega_x(t_f)u'_x(t_f),\,\,
\dot{u}'_y(t_f)\!=\!i\omega_y(t_f)u'_y(t_f).\nonumber\\
\label{final-ux}
\eeqa
The corresponding $G_1(t_f)$ and $G_2(t_f)$ take the form
\beqa
G_1(t_f)&=&c_x a_x(t_f)+c_y a_y(t_f),
\nonumber\\
G_2(t_f)&=&c'_x a_x(t_f)+c'_y a_y(t_f),
\eeqa
where, using Eq. (\ref{final-ux}),
\beqa
c_x&\!\!=\!&-i\sqrt{2\omega_x(t_f)}u_x(t_f),\,
c_y\!\!=\!-i\sqrt{2\omega_y(t_f)}u_y(t_f),\nonumber\\
c'_x&\!\!=\!&-i\sqrt{2\omega_x(t_f)}u'_x(t_f),\,
c'_y\!\!=\!-i\sqrt{2\omega_y(t_f)}u'_y(t_f),\nonumber\\
\label{forms}
\eeqa
which, taking into account that   $I_+(t_f)=n_x(t_f)+n_y(t_f)$,
satisfy the relations
\beq
c^{\ast}_xc_x+c'^{\ast}_xc'_x=1,~~c^{\ast}_yc_y+c'^{\ast}_yc'_y=1,~~c^{\ast}_xc_y+c'^{\ast}_xc'_y=0.
\label{c11}
\eeq
A property of invariants that we shall use repeatedly is that if $|\psi(t)\ra$ is a solution of the Schr\"odinger equation,
then $I(t)|\psi(t)\ra$ is also a solution, as can be readily be checked from the equation satisfied by $I(t)$.
We shall also use that there is an elementary (ground state) solution that begins at $|0,0\ra_i$ and ends at $|0,0\ra_f$.
If we now
take $|1,0\rangle_i=G^{\dag}_1(0)|0,0\rangle_i$ as the initial state,
the final state will be
\beqa
 |\psi(t_f)\rangle\!&=&\!G^\dag_1(t_f)|0,0\rangle_f\nonumber\\
& =&[c^{\ast}_xa^{\dag}_x(t_f)\!+\!c^{\ast}_ya^{\dag}_y(t_f)]|0,0\rangle_f\nonumber\\
\! &=&\!c^{\ast}_x|1,0\rangle_f+c^{\ast}_y|0,1\rangle_f.
\eeqa
Using the normalization condition at final time, we thus find the additional relation
\beq
\langle \psi(t_f)|\psi(t_f)\rangle=c^{\ast}_xc_x+c^{\ast}_y c_y=1,
\label{c12}
\eeq
and similarly we  find, using $G_2$,
\beq
c'^{\ast}_xc'_x+c'^{\ast}_y c'_y=1.
\label{c13}
\eeq
We have found the relations (\ref{c12}) and (\ref{c13}) using for convenience a particular state, but notice that they are general, i.e.,  they are valid as long as $H(t)$ is designed from the invariant $I_+(t)$.
This is because the coefficients do not depend on the state considered but on the Hamiltonian evolution.

%
From the Eqs. (\ref{c11}), (\ref{c12}) and (\ref{c13}),
\beq
c^{\ast}_xc_x=c'^{\ast}_yc'_y,~~c^{\ast}_yc_y=c'^{\ast}_xc'_x.
\label{cmore}
\eeq
Thanks to the relations (\ref{c11}), (\ref{c12}), (\ref{c13}), and (\ref{cmore})
several important results will depend only on the parameter
\beq
b\equiv c^{\ast}_xc_x,
\label{b}
\eeq
so we use a special notation for it.
We shall find shortly that this is exactly the parameter that appears in the expression of the energy increment (\ref{increment}).
\section{Final state\label{fs}}
\subsection{Final state for a given initial eigenstate $|n,k\ra_i$}
Driven by the Hamiltonian designed using $I_0$, the evolution from the ground state $|0,0\rangle_i$ will end at the ground state $|0,0\rangle_f$.
To see the fate of an arbitrary initial eigenstate of $H(0)$, $|n,k\rangle_i$, we construct the invariant
which is initially
\beq
I_{n,k}(0)=\frac{[G^{\dag}_1(0)]^n[G^{\dag}_2(0)]^k}{\sqrt{n!k!}}=\frac{[a^{\dag}_x(0)]^n[a^{\dag}_y(0)]^k}{\sqrt{n!k!}},
\eeq
such that $I_{n,k}(0)|0,0\rangle_i=|n,k\rangle_i$.
At final time it is
\beqa
&&I_{n,k}(t_f)=\frac{1}{\sqrt{n!k!}}[G^{\dag}_1(t_f)]^n[G^{\dag}_2(t_f)]^k\nonumber\\
&&=\frac{1}{\sqrt{n!k!}}[c^{\ast}_xa^{\dag}_x(\!t_f\!)\!+\! c^{\ast}_ya^{\dag}_y(\!t_f\!)]^n[c'^{\ast}_xa^{\dag}_x(\!t_f\!)\!+\! c'^{\ast}_ya^{\dag}_y(\!t_f\!)]^k
\nonumber\\
&&=\frac{1}{\sqrt{n!k!}}\sum^{n}_{i=0}C(n,i)[c^{\ast}_xa^{\dag}_x(t_f)]^{n-i}[c^{\ast}_ya^{\dag}_y(t_f)]^i
\nonumber\\
&&\times\sum^k_{j=0}C(k,j)[c'^{\ast}_xa^{\dag}_x(t_f)]^{k-j}[c'^{\ast}_ya^{\dag}_y(t_f)]^j,
\eeqa
where $C(n,k)\equiv\frac{n!}{k!(n-k)!}$.
The final state will be
\beqa
&&|\psi(t_f)\rangle=I_{n,k}(t_f)|0,0\rangle_f\nonumber\\
&&=\sum^{n}_{i=0}\!\sum^{k}_{j=0}\!\sqrt{\!C(\!n,i\!)C(\!k,j\!)C(\!n\!+\!k\!-\!i\!-\!j,k\!-\!\!j\!)C(\!i\!\!+\!j,i\!)}
\nonumber\\
&&\times\, (c^{\ast}_x)^{ n-i}(c^{\ast}_y)^{i}(c'^{\ast}_x)^{k-j}(c'^{\ast}_y)^j|n+k-i-j,i+j\rangle_f.\nonumber\\
\label{finalstate}
\eeqa
We can now compute the final energy
$\langle \psi(t_f)|H(t_f)|\psi(t_f)\rangle$ for the given initial state $|n,k\rangle_i$
%
to find, for any $n,k$ pair, Eq. (\ref{increment}) with $b=|c_x|^2$, see the details in Appendix \ref{certify}.
%
%
%
%
%
%
%
%
%
\subsection{Perfect state transfers\label{pst}}
In view of Eq. (\ref{increment}), the zeros of $b=c^{\ast}_xc_x$
(as a function of the parameters $\lambda$, $t_f$, and $w_2$) play an important role.
For  $b=0$, the initial and final energy are equal for all $n,k$. Moreover, since
$c_x=c_y'=0$ for $b=0$, see Eq. (\ref{cmore}) (equivalently
$u_x(t_f)=0$ and $u_y'(t_f)=0$),
$c_y$ and $c'_x$ become complex numbers with unit modulus, see Eq. (\ref{c11}),
and with phases $\phi_y$ and $\phi'_x$ that depend on the final point of the trajectories,
see Eq. (\ref{forms}). The linear invariants $G_{1,2}$ at final time become
\beqa
G_1(t_f)|_{b=0}&=&e^{i\phi_y}a_y(t_f),
\nonumber\\
G_2(t_f)|_{b=0}&=&e^{i\phi'_x}a_x(t_f),
\eeqa
so that the final state corresponding to $|n,k\ra_i$ is, for $b=0$,
\beqa
&&\!|\psi(t_f)\rangle|_{b=0}=I_{n,k}(t_f)|0,0\rangle_f\nonumber\\
&&\!=\!(c^{\ast}_y)^n(c'^{\ast}_x)^k|k,n\rangle_f
\!=\!e^{-i(n\phi_y+k\phi'_x)}|k,n\rangle_f.
\label{finals}
\eeqa
In other words, the condition $b=0$ implies  a perfect-transfer  STA protocol.
Notice also the quadratic invariant evolutions
when $b=0$,
\beqa
I_1(0)=n_x(0)\rightarrow I_1(t_f)=n_y(t_f),\\
I_2(0)=n_y(0)\rightarrow I_2(t_f)=n_x(t_f).
\eeqa
%
%
\begin{figure}[tbp]
\scalebox{0.7}[0.7]{\includegraphics{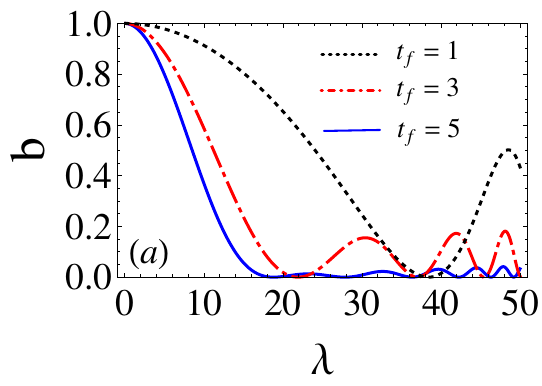}}
\scalebox{0.7}[0.7]{\includegraphics{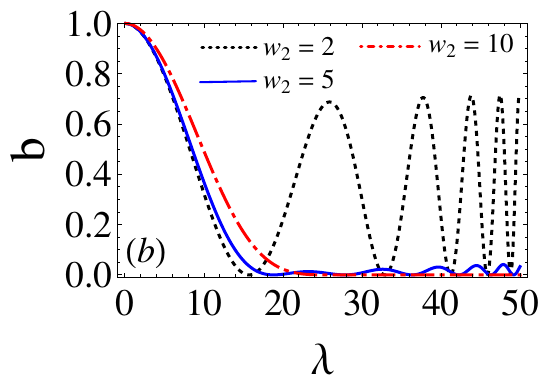}}
\caption{(Color online) $b$ versus $\lambda$ (a) for different final times
and $w_2=5$; (b) for different $w_2$ and $t_f=5$. }\label{figb1}
\end{figure}
As Fig. \ref{figb1} demonstrates, for a given $w_2$ and $t_f$ one can always find values of $\lambda$ that make $b=0$.
Presently the existence of such values is just an observation. Finding a reason why they do indeed occur is  left as an open question.
Figure \ref{figb2} shows the minimum of such $\lambda$ values as a function of $t_f$. While the minimum value increases for shorter times,
the dependence on $w_2$ is not as simple.
We emphasize that $t_f$ is arbitrary so that the process may be clearly non-adiabatic during the transient protocol. Figure
\ref{figh} gives some examples of transient excitations consistent with a vanishing final excitation for two different initial states and two different process times.
%
\begin{figure}[tbp]
\scalebox{0.7}[0.7]{\includegraphics{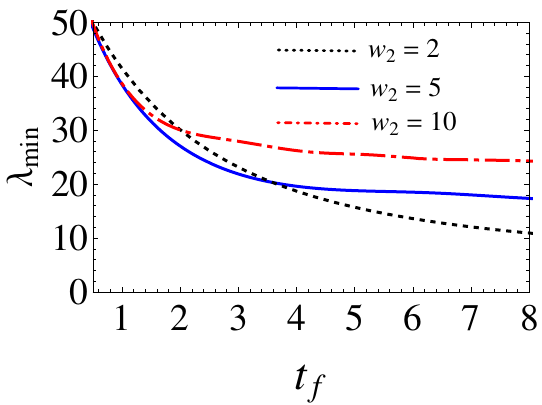}}
\caption{(Color online) For given $t_f$, $\lambda_{min}$ is the minimum $\lambda$ to realize $b=0$.}\label{figb2}
\end{figure}

\begin{figure}[htb]
\scalebox{0.4}[0.4]{\includegraphics{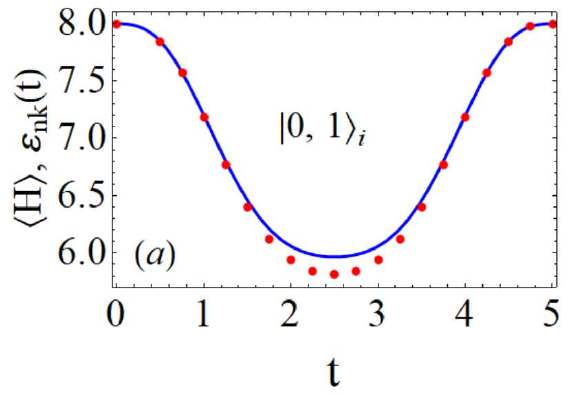}}
\scalebox{0.4}[0.4]{\includegraphics{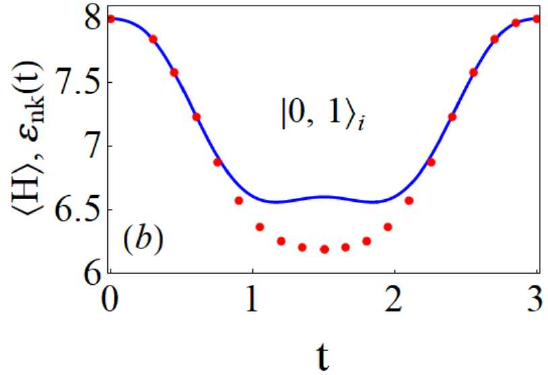}}
\scalebox{0.4}[0.4]{\includegraphics{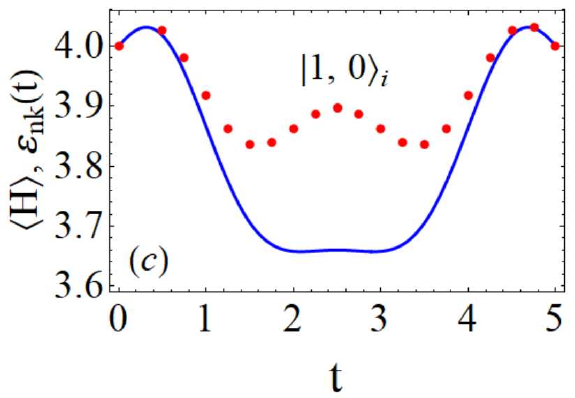}}~~~
\scalebox{0.4}[0.4]{\includegraphics{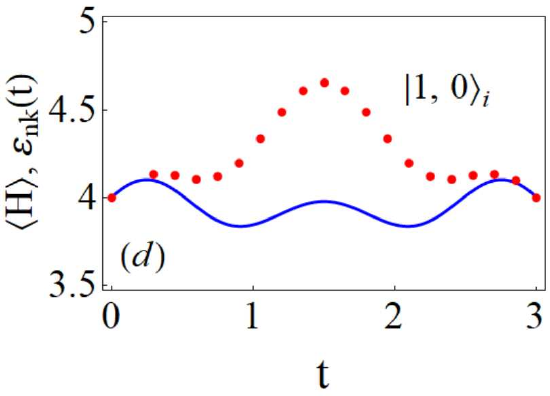}}
\caption{(Color online) Instantaneous eigenenergies  of the Hamiltonian (blue solid lines)
${\cal E}_{nk}(t)\equiv\omega_1(t)(n+1/2)+\omega_2(t)(k+1/2)$,
and the expectation value of the energy (red points) versus time for two different initial states
$|n,k\ra_i$ and  $w_2=5$. For (a) and (c): $t_f=5$, $\lambda=18.81$; for (b) and (d): $t_f$=3, $\lambda=21.89$.}\label{figh}
\end{figure}

\section{Discussion}
We have found that a combination of invariants, the 2D invariant $I_0$ proposed  by Simsek and Mintert \cite{Simsek2021},
and the 2D linear invariants used in Tobalina et al. \cite{Tobalina2020}, allow for fast inverse engineering state transfers,
more specifically vibrational quantum number swapping for any (not necessarily known)
quantum numbers, in coupled oscillators. The linear invariants and their powers
serve to solve the ``degeneracy problem'' of $I_0$, by which its commutation with the Hamiltonian at boundary times does not guarantee perfect transfer
except for the ground state. They also provide explicit expression for the Lewis-Riesenfeld final phase for each state, see Eq. (\ref{finals}).
Different systems may be described  as coupled oscillators, and we have paid special attention to a particle in a trap whose final
form is rotated by $\pi/2$ with respect to the initial one. In this case the method allows for rotations that produce final eigenstates
of the Hamiltonian rotated with respect to the initial eigenstates.
The method can be extended  to arbitrary rotation angles, see Appendix \ref{arbi},
as well as to more complex operations such as a trap translation combined with its rotation, i.e. to the process
considered in  \cite{Simsek2021}. The methodology presented here may be extended into problems in higher dimensions such as ion separations \cite{Simsek2021b}.  A further extension concerns the boundary conditions for the potential.
Suppose that a quantum number exchange is desired, namely $|n,k\ra_i  \to |k,n\ra_f$ up to phase factors, with $M_{12}(0)=M_{12}(t_f)=0$,
as in the main text, but such that the initial and final oscillators
are equal, i. e., such that $(M_{jj})_f=(M_{jj})_i$, $j=1,2$. The process may be designed in two STA steps,
the first one as in the main text, and
the second one as an STA expansion or compression of the oscillators \cite{Chen2010a} keeping the coupling $M_{12}$ zero.
Quite different boundary conditions and Hamiltonian parameter controllability have been set  in ref. \cite{Villazon2019}, where an STA
approach based on counter-diabatic driving and unitary transformations to get rid of terms difficult to implement \cite{Ibanez2012}  was applied, see also ref. \cite{JunJing}.
A detailed comparison will be carried out elsewhere.

Finding perfect transfers requires fine tuning an invariant  parameter $\lambda$. This tuning is done   by calculating  a
complex valued classical trajectory, so that a final boundary condition $(u_x(t_f)=0)$ is satisfied.  The resulting processes
are defined by time dependent functions $\theta(t)$, $\omega_{1,2}(t)$. As the process time is decreased, one of the frequencies for the principal axes may become imaginary, which implies a repeller in that direction. This effect is well known in fast STA
expansion or compression processes \cite{Chen2010a} and may or may not represent a problem to implement it in the laboratory depending on the physical setting. The detailed discussion of specific implementations goes beyond the scope of the present article, but we
point out that in a trapped ion setting, the rotation of the trap with varying frequencies and controlled
angular speed are possible with ``point harmonic traps'' making use of concentric rings to implement a ponderomotive potentials and a rotating electrostatic quadrupole potential \cite{Tobalina2021,Urban2019,UrbanThesis2019}.  Different (two particle) implementations may be based on two ions in controllable double wells \cite{Sagesser2020}.

\section{Acknowledgments}
We are very grateful to Florian Mintert,  Selwyn Simsek, Sof\'\i a Mart\'\i nez Garaot, and Mikel Palmero  for many discussions.
This work was
supported by  National Natural Science Foundation of China (Grant
No. 12104390), the Natural Science Foundation of Henan Province (Grants No. 212300410238),
 and by PGC2018-101355- B-I00 (MCIU/AEI/FEDER,UE).

\appendix
\section{Use of the Wigner function to calculate final energies}\label{sectionwig}
At initial time, the Hamiltonian is given by
\beq
H(0)=\frac{p^2_x}{2}+\frac{p^2_y}{2}+
\frac{1}{2}\omega^2_1(0)x^2+\frac{1}{2}\omega^2_2(0)y^2.
\eeq
Consider as initial states the product of eigenstates of the two harmonic oscillators
with vibrational quantum numbers $n$ and $k$,
\beq
|\Psi(x,y;0)\rangle=|\phi^{(1)}_n(x;0)\rangle|\phi^{(2)}_k(y;0)\rangle
\eeq
The corresponding Wigner function is
$W(x,y,p_x,p_y;0)=W^{(1)}_n(x,p_x;0)W^{(2)}_k(y,p_y;0)$, where
\beqa
W^{(1)}_n(x,p_x;0)&=&\!\int dz \frac{e^{-{ip_x z}}}{2\pi}\phi^{(1)}_n\!\!\left(x\!+\!\frac{z}{2};0\right)\nonumber\\
&\times& \phi^{\ast(1)}_n\!\!\left(x\!-\!\frac{z}{2};0\right),
\nonumber\\
W^{(2)}_k(y,p_y;0)&=&\!\int dz \frac{e^{-{ip_y z}}}{2\pi}\phi^{(2)}_k\!\left(y\!+\!\frac{z}{2};0\right)\nonumber\\
&\times&\!\phi^{\ast(2)}_k\!\left(y\!-\!\frac{z}{2};0\right).
\nonumber
\eeqa
Because the driving is harmonic we may use classical trajectories  to propagate the Wigner function, as well as Liouville's theorem, namely,
$W(x_f,y_f,p_{x,f},p_{y,f}; t_f)=W(x,y,p_x,p_y; 0)$, where $x_f,y_f,p_{x,f},p_{y,f}$ is the  classical trajectory with initial conditions
$x,y,p_x,p_y$ at time $t=0$. Thus in the Weyl-Wigner formulation,
the final energy may be computed easily using the initial Wigner function as weight function and
 classical trajectories to compute the Weyl transform of the energy operator, $H_W(t_f)$, which is simply the classical expression of the energy, as
 $H(t_f)$ does not involve any cross terms between positions and momenta,
\beqa
\langle H(t_f)\rangle&=&\int\!\!\!...\!\!\int dxdydp_{x}dp_{y}W(x,y,p_{x},p_{y};0)
\nonumber\\
&\times&
H_W(x_f,y_f,p_{x,f},p_{y,f};t_f).
\eeqa
The numerical computations proceed by discretizing the integral into
a mesh of initial phase points which serve as initial conditions to calculate
classical trajectories whose final values provide the corresponding final
energies.

\section{Arbitrary final rotation angle\label{arbi}}
In the main text, we consider rotations of the trap from $0$ to $\pi/2$. In fact the use of invariants to design fast state transfer processes
may also be applied to an arbitrary final rotation angle $\gamma$.
Note that for an arbitrary rotation angle $\gamma$
the Cartesian coordinates $x, y$ are still useful as a fixed ``laboratory frame'' for calculations, but
they are no more a privileged frame to represent states and Hamiltonians, as $x$ and $y$ directions will be generally coupled in the final configuration.
Instead, it is convenient to express initial and final states (similarly initial and final number operators, or creation and annihilation operators) in terms of
uncoupled rotated coordinates $q_1, q_2$ and momenta $p_1, p_2$, see Eqs. (\ref{eq2}) to (\ref{Hlab}).
In that representation perfect rotations  take the form $|n_1,k_2\ra_i\to|n_1,k_2\ra_f$, up to phase factors, and
at boundary times $t_b=0,t_f$, $H(t_b)=H_1(t_b)+H_2(t_b)$, where
\beqa
H_1(t_b)&=&\frac{p^2_1}{2}+\frac{1}{2} w^2_1 q_1^2,
\nonumber\\
H_2(t_b)&=&\frac{p^2_2}{2}+\frac{1}{2} w^2_2 q_2^2.
\eeqa
The corresponding invariant is (using the lab frame as in the main text) $I_0(t_b)=\frac{1}{2}X^T \Gamma(t_b)X$, and
\beq
\Gamma\!=\!\!\left(\!\!
  \begin{array}{cc}
    \!\!R^{-2}(t_b)\!\! &\!\! 0\!\! \\
    \!\!0\!\! & \!\!R^2(t_b)\!\! \\
  \end{array}
\!\!\right)\!=\!\left(\!\!
  \begin{array}{cc}
    \!\!M^{1/2}(t_b)\!\! &\!\! 0\! \\
    \!\!0\!\! &\!\! M^{-1/2}(t_b)\!\! \\
  \end{array}
\!\!\right)\!,
\eeq
where now the boundary conditions for $R$ and $M$ at $t_b$ must be adapted to the corresponding potential.
Finally, we get
\beq
I_0(t_b)=n_1(t_b)+n_2(t_b)+1,
\eeq
similarly to the main text, but now the number operators correspond to the two  uncoupled oscillators in the orthogonal principal axes of the initial or final trap.
The auxiliary invariants can be worked out similarly to the main text to achieve perfect state transfers.

%
%
\section{Eq. (\ref{increment}) and meaning of $b$}\label{certify}
The initial state is assumed to be $|\psi(0)\rangle=|n,k\rangle_i$. For any process where $H(t)$ is found from the invariant $I_+(t)$ the  final state is given by Eq. (\ref{finalstate}), where, there are in general several contributions corresponding to a given
state of the (final) basis. It is thus useful to reorder the combination and group them together as
\beq
|\psi(t_f)\rangle=
\sum^{n+k}_{s=0} d_{n+k-s,s}|n+k-s,s\rangle_f
\eeq
where the amplitude for each basis state is
\begin{widetext}
\beq
d_{n+k-s,s}=\sum^{Min(n,s)}_{i=Max(\!0,s\!-\!k\!)}\!\sqrt{\!C(n,i)C(k,s-i)C(n\!+\!k\!-\!s,k\!+i-\!s)C(s,i)}
(c^{\ast}_x)^{ n-i}(c^{\ast}_y)^{i}(c'^{\ast}_x)^{k+i-s}(c'^{\ast}_y)^{s-i}.
\label{amplitude}
\eeq
%
Using Eqs. (\ref{c11}), (\ref{c12}),  and (\ref{c13}),
the corresponding probability $P_{n+k-s,s}=|d_{n+k-s,s}|^2$ is
%
\beqa
P_{n+k-s,s}&=&
\bigg|\sum^{Min(n,s)}_{i=Max(\!0,s\!-\!k\!)}(-1)^i\!\sqrt{\!C(n,i)C(k,s-i)C(n\!+\!k\!-\!s,k\!+i-\!s)C(s,i)}\sqrt{b}^{n+s-2i}\sqrt{1-b}^{k+2i-s}\bigg|^2.\nonumber\\
\label{proba}
\eeqa
%
Let us recall that $b=|c_x|^2=|c_y'|^2$.
Now let us compute the final  energy making use of the fact that the basis functions are eigenstates of $H(t_f)$,
%
\beqa
\langle H(t_f)\rangle&=&\sum_{s=0}^{n+k}P_{n+k-s,s}E_{n+k-s,s}(t_f)
=\!\sum_{s=0}^{n+k}P_{n+k-s,s}\bigg[E_{k,n}(t_f)+(n-s)(w_2-w_1)\bigg]\nonumber\\
&=&\!E_{k,n}(t_f)\!\sum_{s=0}^{n+k}\!P_{n+k-s,s}\!+\!(w_2\!-\!w_1)\sum_{s=0}^{n+k}(n\!-\!s)P_{n+k-s,s}.\nonumber\\
\eeqa
\end{widetext}
where $E_{i,j}(t_f)=(i+1/2)w_2+(j+1/2)w_1$.

This gives Eq. (\ref{increment}) provided
\beqa
\sum_{s=0}^{n+k}P_{n+k-s,s}&=&1,\label{norm-nk}\\
\sum_{s=0}^{n+k}(n-s)P_{n+k-s,s}&=&b(n-k).\label{increment2}
\eeqa
The first relation follows directly from the normalization of the state.
The second one can be worked out ``by hand'' for small $n$ and arbitrary $k$ after some algebra, using Eq. (\ref{proba}), but the
calculation grows rapidly with $n$ because of the combinatorial nature of the problem.
We have thus used Mathematica to check its validity for arbitrary values of $n$ and $k$ (in particular in a loop for all possible
values up to $n=100$, $k=100$).

\bibliography{Bibliography2022}
\bibliographystyle{quantum}


\end{document}